\documentclass[useAMS,usenatbib]{mn2e}
\usepackage{natbib}
\usepackage{color,graphicx} 
\usepackage{amsmath}        
\usepackage{amsfonts}       
\usepackage{amssymb}        
\usepackage{bm}         
\usepackage{framed}         
\usepackage{upgreek}
\usepackage{xspace}
\usepackage{hyperref}
\usepackage{array}
\newcommand{\kb}{FRB~010621\xspace}

\newcommand{\spitlerfrb}{FRB~121102\xspace}

\newcommand{\dmunits}{$\mathrm{cm}^{-3}\,\mathrm{pc}$\xspace}

\author[Keane]{E.F.~Keane$^{1}$ \\ $^{1}$ SKA Organisation, Jodrell
  Bank Observatory, Lower Withington, Macclesfield, Cheshire, SK11
  9DL, UK \\ } \date{\today} \title[RRATs \& FRBs]{Classifying RRATs
  and FRBs}

\begin{document}

\maketitle

\begin{abstract}
 
  In this paper we consider the fact that the simple criterion used to
  label fast radio transient events as either fast radio bursts (FRBs,
  thought to be extragalactic with as yet unknown progenitors) or
  rotating radio transients (RRATs, thought to be Galactic neutron
  stars) is uncertain.
  We identify single pulse events reported in the literature which
  have never been seen to repeat, and which have been labelled as
  RRATs, but are potentially mis-labelled FRBs.
  We examine the probability that such `grey area' events are within
  the Milky Way. 
  The uncertainty in the RRAT/FRB labelling criterion, as well as
  Galactic-latitude dependent reporting bias may be contributing to
  the observed latitude dependence of the FRB rate, in addition to
  effects such as Eddington bias due to scintillation. 

\end{abstract}


\section{Introduction}
Fast radio burst (FRB) is the name given to a handful of sub-second
radio transient signals detected over the last
decade~\citep{kp15}. Because of their many potential uses as tools for
studying extreme physical environments and the Universe up to
redshifts of $\sim2-3$~\citep{mkg+15}, there is a large effort
underway to search for these signals, both in dedicated ongoing
surveys and in archival data~\citep{tsb+13}. FRBs manifest themselves
as millisecond-duration events with dispersion measures of several
hundred \dmunits. No FRB has yet been observed to emit more than one
pulse~\citep{pkb+15}.

Rotating radio transient (RRATs) is the name given to a group of
sporadically pulsing sources~\citep{km11}. The majority of RRATs have
been seen to repeatedly emit detectable pulses, and it seems clear,
from the results of applying pulsar timing
techniques~\citep{mll+06,mlk+09,kkl+11}, that these sources are
Galactic neutron stars with intermittent and/or highly variable pulsar
emission.

Single-instance RRAT pulses are no different in appearance to FRB
pulses. The criterion for deciding whether such pulses should be
classified as FRB or RRAT is comparison of the dispersion measure (DM)
with the predicted Galactic maximum of this parameter. In this paper,
in \S~\ref{sec:ne2001}, we examine the uncertainty in this
classification. Then, in \S~\ref{sec:rrat_or_frb}, we estimate the
probability that various transients are Galactic or extragalactic,
before presenting our concluding thoughts in
\S~\ref{sec:last_section}.

\section{NE2001 Distances}\label{sec:ne2001}
The parameter used in deciding whether a dispersed radio burst is an
FRB, i.e. extragalactic, is $\mathrm{DM}_{\mathrm{MW}}$, the maximum
Galactic DM contribution for the particular line of sight. If the
burst's DM exceeds $\mathrm{DM}_{\mathrm{MW}}$ then it is
extragalactic, an FRB, with an inferred $\sim$Gpc-scale distance. If
the burst's DM is less than $\mathrm{DM}_{\mathrm{MW}}$ then the
source is Galactic, a RRAT, with an inferred $\sim$kpc-scale
distance. In practice the model used has thusfar always been
NE2001~\citep{cl02}; more specifically ``NE2001b'' as per the
labelling of \citet{sch+12}. Here we investigate the uncertainty of
this model, and thus the uncertainty in the RRAT/FRB classification of
dispersed radio bursts which have not been seen to repeat.

\begin{figure}
  \begin{center}
    \includegraphics[scale=0.45,trim = 0mm 0mm 0mm 10mm, clip, angle=0]{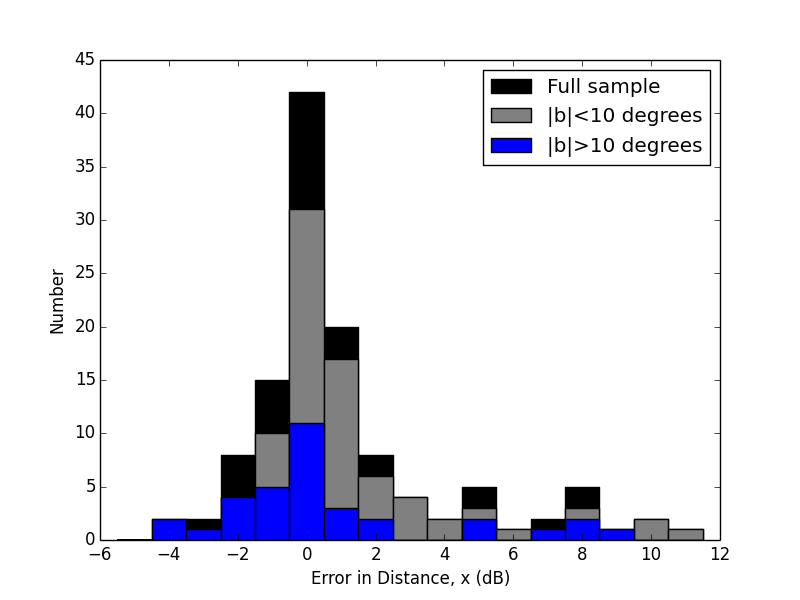}
    \caption{\small{The distribution of the errors in the NE2001
        distance estimate, based on the $120$ radio-detected sources
        in \citet{ver+12}. This distribution also represents (see
        text) the error in
        $x=\mathrm{DM}/\mathrm{DM}_{\mathrm{MW}}$. The error
        distribution is expressed in decibels, such that a negative
        (positive) value corresponds to an underestimate
        (overestimate). The three distributions which are overplotted
        are those based on the entire \citet{ver+12} sample and the
        two sub-samples of sources which are below and above a
        Galactic latitude of
        $10$~degrees.}}\label{fig:ne2001_dist_errors}
  \end{center}
\end{figure}

There are a number of studies in the literature which have addressed
the question of the accuracy of the NE2001 model for predicting
distances. \citet{dtbr09} used distances to $41$ pulsars, determined
from VLBI and pulsar timing measurements of parallax, to estimate the
distribution of distance errors of the model. More recently,
\citet{ver+10,ver+12} identified and corrected the Lutz-Kelker bias in
the by-then enlarged sample of pulsar distance measurements derived
from parallax, supernova remnant associations and neutral hydrogen
absorption. Taking the $120$ sources in \citet{ver+12} with DM
measurements (i.e. those detected at radio wavelengths) we produce a
Lutz-Kelker corrected version of the distance error distribution of
\citet{dtbr09}, shown in Figure~\ref{fig:ne2001_dist_errors}. 

An overestimate (underestimate) of the distance corresponds to an
underestimate (overestimate) in the DM inferred \textit{at the correct
  distance}, and correspondingly an overestimate (underestimate) in
$x=\mathrm{DM}/\mathrm{DM}_{\mathrm{MW}}$. Thus the distribution of
the NE2001 distance errors maps directly to NE2001 $x$ errors,
i.e. RRAT/FRB classification. We can see that the distribution is
skewed, with overestimates more likely with a ratio of $70:50$.
However, there is evidence of latitude dependence in the
distribution. Considering only those pulsars within $10$ degrees of
the Galactic plane the ratio of underestimated:overestimated distances
is $31:55$, but is $19:15$ for sources that are more than $10$ degrees
from the plane. Assuming that finer resolution in Galactic latitude
would see an insufficiently large sample for further resolution, below
we consider only these three distributions in estimating the accuracy
of RRAT/FRB labelling.

\section{FRB or RRAT?}\label{sec:rrat_or_frb}
A search of the literature (see Table~\ref{tab:ne2001_dist_errors} for
references) shows that there are $12$ sources labelled as RRATs, which
have never been seen to show a second radio pulse. In
Table~\ref{tab:ne2001_dist_errors} these are considered in addition to
the 16 published FRBs. The probability that each of these events is
extragalactic is tabulated, for each of the three distributions of
NE2001 errors discussed above: derived from (i) the entire
\citet{ver+12} sample; (ii) the $|b|<10$ degrees sub-sample; and (iii)
the $|b|>10$ degrees sub-sample. Based solely on this simple
assessment of NE2001 uncertainties, FRB classifications seem to be
mostly sound, with perhaps one source being of uncertain progeny: \kb
has a $\sim 70\%$ probability of being extragalactic. Of the `RRAT'
sources, J1354$+$24 appears to have the highest likelihood of being a
mis-labelled FRB, with a probability of $\sim 20 - 40\%$ of being
extragalctic.

Additional outside information can be used to build upon these
estimates, where it is available. For example while one might argue
the lack of scattering in \kb favours an extragalactic
origin~\citep{lkmj13}, identifying a Galactic explanation for \kb has
been the subject of investigation by \citet{kskl12} and \citet{bm14},
the latter suggesting that ionised gas unaccounted for in NE2001 can
explain much or all of the excess dispersion. Another key piece of
additional information is the limit on repeatability. For \spitlerfrb,
\citet{sch+14} suggest the observed spectrum, which is dramatically
positive, is not intrinsic and a result of a far side-lobe
detection. \citet{kon15} suggest the observed spectrum is an effect of
free-free absorption in a nebula surrounding an extragalactic
magnetar. In both scenarios the $\sim 4.5$ hours of followup may be
insufficient to rule out underlying repetition. Obtaining key
additional information like this will be essential to elucidate the
origins of sources like J1354$+$24.

\begin{table*}
  \begin{center}
    \caption{\small{Here we list the $12$ RRAT sources in the
        literature which have not yet been seen to show repeat pulses,
        as well as the 16 published FRBs. The references in the table
        are given in the second column and are: [1] \citet{tsb+13};
        [2] \citet{cha16}; [3] \citet{pbb+15}; [4] \citet{mls+15}; [5]
        \citet{rsj15}; [6] \citet{lbm+07}; [7]; \citet{bb14}; [8]
        \citet{sch+14}; [9] \citet{kkl+11}; [10] \citet{kkl+15}; [11]
        \citet{bb10}; [12] \citet{bbj+11}. For each signal we then
        give the excess DM parameter $x$ in the third column and (in
        dB) in the fourth column. The fifth, sixth and seventh columns
        each give the probability that the signal is an FRB, for the 3
        distributions considered in the text. In each case the exact
        number of sources is given in parentheses next to the
        probability; the error on the probabilities is simply the
        Poisson $95\%$ confidence level~\citep{geh86}. Based on each
        source's Galactic latitude the more relevant distribution out
        of $P_{\mathrm{ii}}$ and $P_{\mathrm{iii}}$ has been
        emboldened. The $10$ FRBs referred in the first row, 
        are: FRBs 110703, 090625, 130729, 110220, 121002, 120127,
        140514, 110626, 110523 and 130626.  Similarly, the $8$ RRATs
        referred to in the final row are: J0441$-$04, J0923$-$31,
        J1311$-$59, J0513$-$04, J1709$-$43, J1649$-$46, J1541$-$42 and
        J0845$-36$.}}\label{tab:ne2001_dist_errors}\setlength{\extrarowheight}{3pt}
    \begin{tabular}{ccccccc}

      \hline \hline 
      Source & Reference & $x$ & $x$ (dB) & $P_{\mathrm{i}}$ & $P_{\mathrm{ii}}$ & $P_{\mathrm{iii}}$ \\
      \hline
      10 further FRBs & [1,2,3,4] & $7.2$ to $34.1$ & $11.5$ to $15.3$ & $>0.99$ (120/120) & $>0.99$ (86/86) & $>0.97$ (34/34) \\
      FRB~131104 & [5] & $10.9$ & $10.4$ & $0.99^{+0.01}_{-0.14}$ (119/120) & $0.99^{+0.01}_{-0.18}$ (85/86) & $\boldsymbol{>0.97}$ (34/34) \\
      FRB~130628 & [2] & $9.0$ & $9.5$  & $0.98^{+0.02}_{-0.15}$ (117/120) & $0.97^{+0.03}_{-0.17}$ (83/86) & $\boldsymbol{>0.97}$ (34/34) \\
      FRB~010724 & [6] & $8.4$ & $9.2$  & $0.97^{+0.03}_{-0.14}$ (116/120) & $0.97^{+0.03}_{-0.17}$ (83/86) & $\boldsymbol{0.97^{+0.03}_{-0.26}}$ (33/34) \\
      FRB~010125 & [7] & $7.2$ & $8.6$  & $0.97^{+0.03}_{-0.14}$ (116/120) & $0.97^{+0.03}_{-0.17}$ (83/86) & $\boldsymbol{0.91^{+0.09}_{-0.36}}$ (31/34) \\
      \spitlerfrb& [8] & $3.0$ & $4.8$  & $0.83^{+0.17}_{-0.08}$ (106/120) & $\boldsymbol{0.90^{+0.10}_{-0.17}}$ (77/86) & $0.85^{+0.15}_{-0.23}$ (29/34) \\
      \kb        & [9] & $1.4$ & $1.5$  & $0.74^{+0.14}_{-0.12}$ (89/120)  & $\boldsymbol{0.73^{+0.27}_{-0.14}}$ (63/86) & $0.77^{+0.23}_{-0.24}$ (26/34) \\
      J1354$+$24 & [10] & $0.9$ & $-0.4$  & $0.25^{+0.09}_{-0.07}$ (30/120)  & $0.20^{+0.10}_{-0.07}$ (17/86) & $\boldsymbol{0.38^{+0.24}_{-0.14}}$ (13/34) \\
      J1610$-$17 & [11] & $0.7$ & $-1.7$  & $0.07^{+0.04}_{-0.04}$ (8/120)   & $0.02^{+0.05}_{-0.01}$ (2/86) & $\boldsymbol{0.18^{+0.17}_{-0.09}}$ (6/34) \\
      J1135$-$49 & [12] & $0.7$ & $-1.8$  & $0.07^{+0.04}_{-0.04}$ (8/120)   & $0.02^{+0.05}_{-0.01}$ (2/86) & $\boldsymbol{0.18^{+0.17}_{-0.09}}$ (6/34) \\
      J1059$-$01 & [10] & $0.5$ & $-3.0$  & $0.03^{+0.05}_{-0.02}$ (4/120)   & $0.01^{+0.04}_{-0.01}$ (1/86) & $\boldsymbol{0.09^{+0.15}_{-0.06}}$ (3/34) \\
      8 further RRATs & [9,10,11,12] & $-0.4$ to $0.1$ & $-4.1$ to $-10.7$ & $<0.02$ (0/120) & $<0.03$ (0/86) & $<0.09$ (0/34) \\

    \end{tabular}
  \end{center}
\end{table*}

\section{Discussion \& Conclusions}\label{sec:last_section}
We have considered the uncertainty in the DM excess parameter
$x=\mathrm{DM}/\mathrm{DM}_{\mathrm{MW}}$, crucial in RRAT/FRB
classifications in high time resolution radio surveys of the sky.  By
considering a simple probability-based estimator, we have determined
that the FRB identifications made so far are reasonably secure with
perhaps one exception (\kb). On the other side of the clasification,
we find the `RRAT' J1354$+$24 has a $\sim 20 - 40 \%$ chance of being
a mis-labelled FRB; this suggest it merits a deep observational study
to address the question of whether it repeats. Similarly, the
cumulative probability of at least one RRAT being mis-labelled is at
the $\sim 30 - 80 \%$ level; a dedicated campaign to search for a
second pulse from each of these sources is merited.

A further concern is a potential bias in the reported single pulse
events. In pulsar and transient surveys it is standard practice to
consider a source to be a \textit{bona fide} astrophysical source only
when it has been observed on at least two epochs~\citep{lk05}. This
rule, coupled with finite available observing time for re-observing
pulsar and transient candidates, continues to contribute to the
difficulty in identifying those intermittent pulsars and RRATs with
the longest repetition timescales~\citep{klo+06}. However, this rule
is not applied to FRBs: they are simply reported when detected. Given
this, 
and that much more observing time is spent closer to the Galactic
plane~\citep{bb14} we suspect a Galactic-latitude dependent bias in
the reported events.
To enable correction for this effect survey teams should report single
pulse events which appear to be of astrophysical origin, regardless of
whether they are seen on a second epoch or have
$x<1$. Probability-based classifiers could then be used for
determining the Galactic/extragalactic nature.

The observed FRB rate at Parkes~\citep{psj+14} is higher at higher
latitudes. With a perfect electron density model and no reporting
bias, such a dependence can still arise, as a result of Eddington bias
of the population due to diffractive scintillation at high
latitudes~\citep{mj15}. In this scenario the true rate is that at low
latitudes, and the `boost factor' depends directly on the $\log N -
\log S$ distribution --- the steeper the distribution the larger the
boost. But it is clear that the difference between the observed and
intrinsic rates, and hence in our ability to uncover the true $\log N
- \log S$ and luminosity distributions, is a combination of a number
of effects. In this paper we have tried to tackle the effect of the
uncertainty of NE2001. With this, and a clear picture of single epoch
pulse events, including those with $x<1$ (most, but perhaps not all,
will be within our Galaxy), we will be able to determine trustworthy
metrics for determining the cosmic history of FRBs, whether or not
they are standard candles, and their general utility as tools for
precision cosmology.

\section*{Acknowledgements}
The author would like to thank the referee, J.-P. Macquart, for
valuable input that improved the quality of this paper.
%
%


\begin{thebibliography}{}
  
\bibitem[\protect\citeauthoryear{{Bannister} \& {Madsen}}{{Bannister}
    \& {Madsen}}{2014}]{bm14} {Bannister} K.~W., {Madsen} G.~J., 2014,
  MNRAS, 440, 353

\bibitem[\protect\citeauthoryear{{Burke-Spolaor} \&
    {Bailes}}{{Burke-Spolaor} \& {Bailes}}{2010}]{bb10}
  {Burke-Spolaor}, S., {Bailes} M., 2010, MNRAS, 402, 855

\bibitem[\protect\citeauthoryear{{Burke-Spolaor}
    et~al.}{{Burke-Spolaor} et~al.}{2011}]{bbj+11} {Burke-Spolaor} S.,
  et al., 2011, MNRAS, 416, 2465

\bibitem[\protect\citeauthoryear{{Burke-Spolaor} \&
    {Bannister}}{{Burke-Spolaor} \& {Bannister}}{2014}]{bb14}
  {Burke-Spolaor}, S., {Bannister} K.~W., 2014, ApJ, 792, 19

\bibitem[\protect\citeauthoryear{{Champion} et~al.}{{Champion}
    et~al.}{2016}]{cha16} {Champion} D.~J., et~al., 2016, MNRAS,
  in press, (astro-ph/1511.07746)

\bibitem[\protect\citeauthoryear{{Cordes} \& {Lazio}}{{Cordes} \&
    {Lazio}}{2002}]{cl02} {Cordes} J.~M., {Lazio} T.~J.~W., 2002
  (astro-ph/0207156)

\bibitem[\protect\citeauthoryear{{Deller} et~al.}{{Deller}
    et~al.}{2009}]{dtbr09} {Deller} A.~T., {Tingay}, S.~J., {Bailes},
  M, {Reynolds}, J.~E., 2009, ApJ, 701, 1243

\bibitem[\protect\citeauthoryear{{Gehrels}}{{Gehrels}}{1986}]{geh86} {Gehrels},
  N., 1986, ApJ, 303, 336

\bibitem[\protect\citeauthoryear{{Karako-Argaman}
    et~al.}{{Karako-Argaman} et~al.}{2015}]{kkl+15} {Karakao-Argaman}
  C., et al., 2015, ApJ, 809, 67

\bibitem[\protect\citeauthoryear{{Keane} et~al.}{{Keane}
    et~al.}{2011}]{kkl+11} {Keane} E.~F., {Kramer} M., {Lyne} A.~G.,
  {Stappers} B.~W., {McLaughlin} M.~A., 2011, MNRAS, 415, 3065

\bibitem[\protect\citeauthoryear{{Keane} \& {McLaughlin}}{{Keane} \&
    {McLaughlin}}{2011}]{km11} {Keane} E.~F., {McLaughlin} M.~A.,
  2011, Bulletin of the Astronomical Society of India, 39, 333

\bibitem[\protect\citeauthoryear{{Keane} et~al.}{{Keane}
    et~al.}{2012}]{kskl12} {Keane} E.~F., {Stappers} B.~W., {Kramer}
  M., {Lyne} A.~G., 2012, MNRAS, 425, L71

\bibitem[\protect\citeauthoryear{{Keane} \& {Petroff}}{{Keane} \&
    {Petroff}}{2015}]{kp15} {Keane} E.~F., {Petroff} E., 2015, MNRAS,
  447, 2852

\bibitem[\protect\citeauthoryear{{Kramer} et~al.}{{Kramer}
    et~al.}{2006}]{klo+06} {Kramer} M., {Lyne} A.~G., {O'Brien}
  J.~T., {Jordan} C.~A., {Lorimer} D.~R., 2006, Science, 312, 549

\bibitem[\protect\citeauthoryear{{Kulkarni} et~al.}{{Kulkarni}
    et~al.}{2015}]{kon15} {Kulkarni} S.~R., {Ofek} E.~O., {Neill}
  J.~D., 2015, preprint (astro-ph/1511.09137).

\bibitem[\protect\citeauthoryear{{Lorimer} et~al.}{{Lorimer}
    et~al.}{2007}]{lbm+07} {Lorimer} D.~R., {Bailes} M., {McLaughlin}
  M.~A., {Narkevic} D.~J., {Crawford} F., 2007, Science, 318, 777

\bibitem[\protect\citeauthoryear{{Lorimer} et~al.}{{Lorimer}
    et~al.}{2013}]{lkmj13} {Lorimer} D.~R., {Karastergiou} A., {McLaughlin}
  M.~A., {Johnston} S., 2013, MNRAS, 436, L5

\bibitem[\protect\citeauthoryear{{Lorimer} \& {Kramer}}{{Lorimer} \&
    {Kramer}}{2005}]{lk05} {Lorimer} D.~R., {Kramer} M., 2005,
  Cambridge University Press

\bibitem[\protect\citeauthoryear{Macquart et~al.}{Macquart
    et~al.}{2015}]{mkg+15} Macquart J.-P., Keane, E. F., Grainge, K.,
  et al.\ 2015, ``Fast transients at cosmological distances'', in
  proc. {\em Advancing Astrophysics with the Square Kilometre Array},
  PoS(AASKA14)055

\bibitem[\protect\citeauthoryear{{Macquart} \& {Johnston}}{{Macquart}
    \& {Johnston}}{2015}]{mj15} {Macquart} J.-P., {Johnston} S., 2015,
  MNRAS, 451, 3278

\bibitem[\protect\citeauthoryear{{Masui} et~al.}{{Masui}
    et~al.}{2015}]{mls+15} {Masui} K., et~al., 2015, Nature, 528, 523

\bibitem[\protect\citeauthoryear{{McLaughlin} et~al.}{{McLaughlin}
    et~al.}{2006}]{mll+06} {McLaughlin} M.~A. et~al., 2006, Nature,
  493, 817

\bibitem[\protect\citeauthoryear{{McLaughlin} et~al.}{{McLaughlin}
    et~al.}{2009}]{mlk+09} {McLaughlin} M.~A. et~al., 2009, MNRAS, 400,
  1431

\bibitem[\protect\citeauthoryear{{Petroff} et~al.}{{Petroff}
    et~al.}{2014a}]{psj+14} {Petroff} E. et~al., 2014, ApJ, 789, L26

\bibitem[\protect\citeauthoryear{{Petroff} et~al.}{{Petroff}
    et~al.}{2015a}]{pbb+15} {Petroff} E. et~al., 2015a, MNRAS, 447,
  246

\bibitem[\protect\citeauthoryear{{Petroff} et~al.}{{Petroff}
    et~al.}{2015b}]{pkb+15} {Petroff} E. et~al., 2015b, MNRAS, 451, 3922

\bibitem[\protect\citeauthoryear{{Ravi} et~al.}{{Ravi}
    et~al.}{2015}]{rsj15} {Ravi} V., {Shannon}, M.~R., {Jameson}, A.,
  2015, ApJ, 799, L5


\bibitem[\protect\citeauthoryear{{Schnitzeler} et~al.}{{Schnitzeler}
    et~al.}{2012}]{sch+12} {Schnitzeler} D. et~al., 2012, MNRAS, 427,
  664

\bibitem[\protect\citeauthoryear{{Spitler} et~al.}{{Spitler}
    et~al.}{2014}]{sch+14} {Spitler} L.~G. et~al., 2014, ApJ, 790,
  101

\bibitem[\protect\citeauthoryear{{Thornton} et~al.}{{Thornton}
    et~al.}{2013}]{tsb+13} {Thornton} D. et~al., 2013, Science, 341,
  53

\bibitem[\protect\citeauthoryear{{Verbiest} et~al.}{{Verbiest}
    et~al.}{2010}]{ver+10} {Verbiest} J. et~al., 2010, MNRAS, 405, 564

\bibitem[\protect\citeauthoryear{{Verbiest} et~al.}{{Verbiest}
    et~al.}{2012}]{ver+12} {Verbiest} J. et~al., 2012, ApJ, 755, 39


\end{thebibliography}
\bibliographystyle{mnras}

\end{document}